\documentclass[fleqn,10pt,a4paper]{article}
\usepackage{amsmath,amssymb,bm}
\usepackage{cite,epsfig,color,graphicx}

\newcommand{\xd}{\dot{x}}
\newcommand{\qm}{\frac{q}{m}}
\newcommand{\xdd}{\ddot{x}}
\newcommand{\xddd}{\dddot{x}}

\title{Aspects of electromagnetic radiation reaction in strong fields}

\author{David A. Burton\thanks{Department of Physics,
Lancaster University, Lancaster, LA1 4YB, UK
and Cockcroft Institute, Daresbury, WA4 4AD, UK.}
\and
Adam Noble\thanks{Department of Physics,
SUPA and University of Strathclyde, Glasgow, G4 0NG, UK.}
}

\date{January 18, 2014}

\begin{document}
\maketitle
\begin{abstract}
With the recent advances in laser technology, experimental investigation of radiation reaction phenomena is at last becoming a realistic prospect. A pedagogical introduction to electromagnetic radiation reaction is given with the emphasis on matter driven by ultra-intense lasers. Single-particle, multi-particle, classical and quantum aspects are all addressed.
\end{abstract}

\section{Introduction}
The response of matter to its own electromagnetic radiation has captivated physicists ever since the late 19th century when Lorentz undertook his ambitious programme to account for all macroscopic electrodynamical and optical phenomena in terms of the microscopic behaviour of electrons and ions. The reason for this fascination is easy to appreciate when one considers the pivotal roles that electricity and magnetism have played in the astonishing pace of technological advancement over the last century.  In this context it is even more remarkable that, to the present day, debate continues to rage unabated about the dynamical behaviour of an electron, the simplest elementary particle, when it is driven by ultra-high strength electromagnetic fields. The next generation of ultra-high field facilities, such as ELI~\cite{ELI}, will provide lasers with field intensities of order $10^{23}\,{\rm W\,cm^{-2}}$ and, for the first time, will offer the opportunity to explore the behaviour of matter when the force due to the electron's own emission exceeds the force due to an applied laser field. Such considerations are also of vital importance for exploring novel sources of coherent electromagnetic pulses of zeptosecond duration~\cite{kaplan:2002}.

The purpose of this article is to introduce the reader to electromagnetic radiation reaction in the context of ultra-intense laser-matter interactions. Much effort has been devoted to { the subject of radiation reaction} over many decades by numerous researchers, and it would be impossible to include everything in a single pedagogical article. Instead, our aim is to give the reader a flavour of the contemporary issues in this field and encourage them to pursue the technical details elsewhere. { For example, Erber~\cite{erber:1961} has given an overview of the development of classical radiation reaction up to the beginning of the 1960s, and the seminal book by Rohrlich~\cite{rohrlich:2007} also offers a historical perspective.} 

In parallel with the motivation provided by the advent of ELI and other planned large-scale facilities, it is worth noting that radiation reaction in strong fields has also received much attention from the gravitational physics community during recent years~\cite{poisson:2011}. Understanding the behaviour of inspiralling black hole binary systems requires efficient and accurate numerical methods for modelling strong-field {\it gravitational} radiation reaction, and such work is vital for the development of matched filters (templates) used to extract information from a binary system's gravitational wave emission. Some of the recent progress in electromagnetic radiation reaction~\cite{gralla:2009} has been made as a consequence of modern interest in the gravitational radiation reaction of extreme-mass-ratio binary systems.

\section{Non-relativistic considerations}
Ultra-intense laser-matter interactions are highly relativistic. However, an appreciation of why the subject of radiation reaction has enticed so many researchers is simplest to gain by considering how a {\it non-relativistic} particle responds to its own radiation. 
\subsection{Abraham-Lorentz equation}
Newton's equation of motion
\begin{equation}
\label{lorentz_force}
m {\bm a} = q({\bm E}_{\rm ext} + {\bm v}\times {\bm B}_{\rm ext})
\end{equation}
describes the response of a non-relativistic {\it point} particle, with mass $m$ and charge $q$, to an {\it external} electric field ${\bm E}_{\rm ext}$ and {\it external} magnetic field ${\bm B}_{\rm ext}$. However, an accelerating charged particle emits electromagnetic radiation, and the energy carried away by the radiation must be accounted for. Simply adding the particle's own contributions ${\bm E}_{\rm self}$, ${\bm B}_{\rm self}$ to the external electric and magnetic fields is 
problematic because the Coulomb field of a point particle diverges at the particle's location. Sophisticated regularization procedures do make this possible \cite{teitelboim:1971,barut:1974}, and yield the same equations as the following.
The Lorentz force exerted by the external fields is augmented with a term ${\bm F}_{\rm rad}$, 
\begin{equation}
m {\bm a} = q({\bm E}_{\rm ext} + {\bm v}\times {\bm B}_{\rm ext}) + {\bm F}_{\rm rad},
\end{equation}
and the form of the radiation reaction force ${\bm F}_{\rm rad}$ may be motivated using a simple argument~\cite{jackson:1999} as follows.

The classical expression for the instantaneous electromagnetic power emitted by a non-relativistic particle is given by the Larmor formula 
\begin{equation}
P = m\tau |\bm{a}|^2
\end{equation}
where $\tau = q^2/6\pi \varepsilon_0 m c^3$ in MKS units and, in particular, $\tau = 2 r_e/ 3 c \sim 10^{-23}~{\rm s}$ for an electron (with $r_e$ the classical electron radius). The work done by the radiation reaction force ${\bm F}_{\rm rad}$ over the time interval $t_1 < t < t_2$ and the energy radiated by the particle over that time interval must balance, and it follows
\begin{equation}
\label{larmor_energy_balance}
\int^{t_2}_{t_1} {\bm F}_{\rm rad} \cdot {\bm v}\, dt = - \int^{t_2}_{t_1} P\, dt.  
\end{equation}
Hence, rewriting the right-hand side of (\ref{larmor_energy_balance}) using an integration by parts leads to
\begin{equation}
\label{larmor_energy_balance_parts}
\int^{t_2}_{t_1} {\bm F}_{\rm rad} \cdot {\bm v}\, dt = \big[ - m\tau {\bm a}\cdot {\bm v}\big]^{t_2}_{t_1} + m \tau \int^{t_2}_{t_1} \frac{d{\bm a}}{dt}\cdot {\bm v}\, dt
\end{equation}
If the boundary term in (\ref{larmor_energy_balance_parts}) vanishes (e.g. the motion is periodic) then
\begin{equation}
\int^{t_2}_{t_1} ({\bm F}_{\rm rad} - m\tau \frac{d{\bm a}}{dt}) \cdot {\bm v}\, dt = 0
\end{equation}
follows immediately, and it is reasonable to identify the radiation reaction force as 
\mbox{${\bm F}_{\rm rad} = m\tau d{\bm a}/dt$.}

The result
\begin{equation}
\label{abraham-lorentz}
m {\bm a} = q({\bm E}_{\rm ext} + {\bm v}\times {\bm B}_{\rm ext}) + m\tau \frac{d{\bm a}}{dt}
\end{equation}
is known as the {\it Abraham-Lorentz equation} and it has the curious property of being a {\it third} order ordinary differential equation for the position of the particle.
More rigorous derivations also lead to (\ref{abraham-lorentz}). 
We are confronted with the unfamiliar situation in which the instantaneous position, velocity {\it and acceleration} of the particle must be specified in order to to find its location at later times.

Equation (\ref{abraham-lorentz}) is solved by ${\bm a}(t) = {\bm a}(0)\exp(t/\tau)$ when the external fields vanish, and this type of solution is known as a {\it runaway} because the particle's acceleration increases exponentially in time unless its initial acceleration ${\bm a}(0)$ is zero. 
Consideration of the time constant $\tau$ for an electron ($\tau \sim 10^{-23}~{\rm s}$) immediately shows that runaway solutions are totally unphysical; an electron at rest would respond to an external perturbation by accelerating to extraordinarily high speeds over very short time scales.
Such behaviour is generic, and not a peculiarity of vanishing external fields: a runaway term can always be added to any solution of (\ref{abraham-lorentz}), and its rapid growth implies it will dominate any acceleration generated by the applied fields.
Runaway solutions must be excluded on physical grounds.

Further investigation of (\ref{abraham-lorentz}) yields more curiosities~\cite{hammond:2010a, hammond:2010b}. Equation (\ref{abraham-lorentz}) may be rewritten as 
\begin{equation}
\label{accel_integral}
m{\bm a}(t) = m{\bm a}(t_0)e^{(t-t_0)/\tau}-\frac{e^{t/\tau}}{\tau}\int^t_{t_0}{\bm F}_{\rm ext}(t^\prime)e^{-t^\prime/\tau}\,dt^\prime
\end{equation}
where $t_0$ is a constant and ${\bm F}_{\rm ext} = q({\bm E}_{\rm ext} + {\bm v}\times {\bm B}_{\rm ext})$. 
Runaway behaviour can be eliminated by demanding that the acceleration tends to zero in the asymptotic future, once all forces have finished acting. This can be implemented in (\ref{accel_integral}) by taking the limits $t_0\rightarrow \infty$, ${\bm a}(t_0)\rightarrow 0$.
Applying the change of variable $s=(t^\prime - t)/\tau$ yields the expression 
\begin{equation}
\label{accel_integral_s}
m{\bm a}(t) = \int^\infty_0 {\bm F}_{\rm ext}(t + \tau s)\,e^{-s}\, ds.
\end{equation}

However, (\ref{accel_integral_s}) exhibits an unphysical phenomenon called {\it pre-acceleration}. The acceleration at time $t$ depends on the applied force {\em at all subsequent times}. The removal of runaway solutions thus requires that the particle is prescient. Inspection of the integrand in (\ref{accel_integral_s}) reveals that the influence of the future force on the present acceleration is exponentially weak: in contrast to the runaways, the smallness of $\tau$ renders pre-acceleration more palatable rather than less.


Although pre-acceleration may be eliminated and causality restored if $t_0=-\infty$ is chosen in (\ref{accel_integral}) instead of $t_0=\infty$, runaway behaviour re-emerges.
\subsection{Non-relativistic Landau-Lifshitz equation}
Runaways and pre-acceleration can be removed simultaneously by reducing the order of the Abraham-Lorentz equation (\ref{abraham-lorentz}). Substituting $m{\bm a} = {\bm F}_{\rm ext} + {\cal O}(\tau)$ into the right-hand side of (\ref{abraham-lorentz}) yields
\begin{align}
\notag
m{\bm a} &= {\bm F}_{\rm ext} + \tau\frac{d{\bm F}_{\rm ext}}{dt} + {\cal O}(\tau^2)\\
\label{abraham_lorentz_iterated}
&= q({\bm E}_{\rm ext}+\bm{v}\times{\bm B}_{\rm ext}) + \tau \bigg[q\Big(\frac{d\bm{E}_{\rm ext}}{dt}+\bm{v}\times\frac{d\bm{B}_{\rm ext}}{dt}\Big) + \frac{q}{m}{\bm F}_{\rm ext}\times{\bm B}_{\rm ext}\bigg] + {\cal O}(\tau^2)
\end{align}
and the {\it non-relativistic Landau-Lifshitz} equation~\cite{landau:1987}
\begin{equation}
\label{non-rel_landau_lifshitz}
m{\bm a} = q({\bm E}_{\rm ext}+\bm{v}\times{\bm B}_{\rm ext}) + \tau \bigg[q\Big(\frac{d\bm{E}_{\rm ext}}{dt}+\bm{v}\times\frac{d\bm{B}_{\rm ext}}{dt}\Big) + \frac{q}{m}{\bm F}_{\rm ext}\times{\bm B}_{\rm ext}\bigg]
\end{equation}
is obtained by dropping ${\cal O}(\tau^2)$ terms. The 
total derivative of the fields is given by $d{\bm{E}}_{\rm ext}/dt = \partial_t \bm{E}_{\rm ext} + ({\bm v} \cdot {\bm \nabla}) \bm{E}_{\rm ext}$, $d{\bm{B}}_{\rm ext}/dt = \partial_t \bm{B}_{\rm ext} + ({\bm v} \cdot {\bm \nabla}) \bm{B}_{\rm ext}$.

Clearly, (\ref{non-rel_landau_lifshitz}) is of the form $m{\bm a} = {\bm F}({\bm x}, {\bm v}, t)$ and it does not suffer from the same pathologies as the Abraham-Lorentz equation. Therefore, (\ref{non-rel_landau_lifshitz}) is generally accepted to be the correct classical equation of motion for a non-relativistic charged point particle if the external fields are sufficiently weak and vary sufficiently slowly (in position and time).
\subsection{Relativistic considerations}
The {\it Lorentz-Dirac equation}~\cite{dirac:1938} (also called the {\it Lorentz-Abraham-Dirac}, or {\it Abraham-Lorentz-Dirac}, equation) is a fully relativistic classical description of a structureless point particle in an applied electromagnetic field $F^{\rm ext}_{ab}$ and has the form
\begin{equation}
\label{Lorentz-Dirac}
m\xdd^a = - q F_{\rm ext}^{ab}\,\xd_b + m\tau \Delta^a{ }_b \xddd^b
\end{equation}
with $q$ the particle's charge, $m$ the particle's rest mass and $\tau = q^2/6\pi m$ in Heaviside-Lorentz units with $c=\epsilon_0=\mu_0=1$. An overdot denotes differentiation with respect to proper time $\lambda$, and the tensor $\Delta^a{ }_b$ is
\begin{equation}
\Delta^a{ }_b = \delta^a_b + \xd^a\xd_b.
\end{equation}
The Einstein summation convention is used throughout, indices are raised and lowered using the metric tensor $\eta_{ab} = {\rm diag}(-1,1,1,1)$ and lowercase Latin indices range over $0,1,2,3$. The particle's $4$-velocity $\xd^a=dx^a/d\lambda$ is normalized as follows:
\begin{equation}
\label{proper_time}
\xd^a\xd_a = -1.
\end{equation}
Equation (\ref{Lorentz-Dirac}) is the relativistic generalization of (\ref{abraham-lorentz}) and exhibits the same pathologies.

Equation (\ref{Lorentz-Dirac}) may be obtained by appealing to the conservation condition
\begin{equation}
\label{conservation_of_energy-momentum}
\partial_a T^{ab} = 0
\end{equation}
satisfied by the stress-energy-momentum tensor $T^{ab}$,
\begin{equation}
T^{ab} = F^{ac} F^b{ }_c - \frac{1}{4}\eta^{ab} F_{cd} F^{cd},
\end{equation}
for the {\it total} electromagnetic field $F_{ab} = F^{\rm self}_{ab} + F^{\rm ext}_{ab}$, where the contribution $F^{\rm self}_{ab}$ is the Li\'enard-Wiechert field of the point particle~\cite{jackson:1999}. The details of the passage from (\ref{conservation_of_energy-momentum}) to (\ref{Lorentz-Dirac}) are rather involved and subtle~\cite{rohrlich:2007}, and will not be addressed here (a recent detailed discussion of the derivation of (\ref{Lorentz-Dirac}) may be found in Ref.~\cite{ferris:2011}).

Perhaps the most notable aspect of Dirac's derivation~\cite{dirac:1938} of (\ref{Lorentz-Dirac}) is the need to renormalize the mass of the electron. Although the concept of renormalization in physics is more usually associated with quantum, rather than classical, electrodynamics, from a historical perspective it is worth noting that Dirac obtained (\ref{Lorentz-Dirac}) a decade before the divergences inherent in loop Feynman diagrams in QED were overcome using renormalization.

The force on the electron contains a term proportional to $d^2 x^a/d\lambda^2$ that leads to an electromagnetic contribution to the electron's mass. However, the shift in mass is infinite for a point electron (since $F^{\rm self}_{ab}$ diverges at the electron's world line) and the bare (unrenormalized) mass must be negatively infinite to yield a finite result for the mass of the electron.

The pathologies inherent in the Lorentz-Dirac equation are removed using an iteration procedure analogous to that used in the passage from (\ref{abraham-lorentz}) to (\ref{non-rel_landau_lifshitz}). The third-order terms in (\ref{Lorentz-Dirac}) (the radiation reaction force) are replaced with the derivative of the first term on the right-hand side of (\ref{Lorentz-Dirac}) (the Lorentz force due to the external field) yielding a second order differential equation for the particle's world line. This procedure is justifiable if the radiation reaction force is a small perturbation to the Lorentz force due to the applied field, and it yields the relativistic Landau-Lifshitz equation~\cite{landau:1987}:
\begin{align}
\label{Landau-Lifshitz}
m\xdd^a = -q F^{ab}_{\rm ext}\,\xd_b - \tau q\, \partial_c F^{ab}_{\rm ext} \xd_b\xd^c
+ \tau \frac{q^2}{m} \Delta^a{ }_b F^{bc}_{\rm ext} F^{\rm ext}_{cd} \xd^d.
\end{align}
The Landau-Lifshitz equation (\ref{Landau-Lifshitz}) is the accepted description of the dynamics of a charged particle when the external field is sufficiently weak and slowly varying in space and time. However, as noted earlier, the fields in forthcoming ultra-high intensity laser facilities will be so strong that the forces due to an electron's emission exceed the Lorentz force on the electron due to the laser pulse. The opportunity to experimentally probe (\ref{Landau-Lifshitz}) is expected to be available during the coming decade, and this has reinvigorated interest in alternatives to (\ref{Landau-Lifshitz}) arising from quantum and more general considerations.
\section{Alternative theories}

Given the difficulties facing the Lorentz-Dirac equation, a number of researchers have proposed alternative theories to describe the response of a particle to its emission of radiation. Although none of these has achieved widespread acceptance, it is of interest to explore the various motivations that led to a number of them, along with their respective advantages and pitfalls. The following is by no means a comprehensive list of theories, but captures a flavour of the various approaches taken.

\subsection{Eliezer--Ford--O'Connell equation}

Perhaps the first attempt to address the deficiencies of the Lorentz-Dirac equation was put forward by Eliezer, a student of Dirac's, in 1948 \cite{Eliezer}. Noting that the equation of motion of a nonrelativistic {\em extended} electron of radius $R$ can very generally be expanded as
\begin{equation}
m{\bm a}- m\tau \frac{d\bm a}{dt}+ \sum^\infty_{n=0} c_n R^n \frac{d^n {\bm a}}{dt^n} = {\bm F}_{\rm ext},
\end{equation}
where 
the $c_n$ are coefficients that depend on the structure of the particle, Eliezer asked the question, can the radius and charge density of the electron be such that
\begin{equation}
m\sum^\infty_{n=0} \big(-\tau \frac{d}{dt}\big)^n {\bm a}= m\big(1+\tau \frac{d}{dt}\big)^{-1} {\bm a}={\bm F}_{\rm ext}.
\end{equation}
Then for such a particle, the equation of motion could be recast as
\begin{equation}
\label{Eq:Eliezer-Ford-O'Connell-nonrelativistic}
m{\bm a}= {\bm F}_{\rm ext}+ \tau \frac{d\bm F_{\rm ext}}{dt},
\end{equation}
which can be made relativistically covariant,
\begin{equation}
\label{Eq:Eliezer-Ford-O'Connell}
m\xdd^a= f^a_{\rm ext}+ \tau\Delta^a{}_b \dot{f}^b_{\rm ext}
\end{equation}
where $\xd^a = dx^a/d\lambda$.

Eliezer made no attempt to interpret the structure of the particle for which (\ref{Eq:Eliezer-Ford-O'Connell}) is the equation of motion, and essentially took it as his starting point. However, some four decades later Ford and O'Connell \cite{FO1,FO2} rediscovered (\ref{Eq:Eliezer-Ford-O'Connell-nonrelativistic}) as the classical limit of a quantum { Langevin equation for extended electrons, a unified description of radiation reaction and (quantum and thermal) fluctuations} (see Ref.~\cite{oconnell:2012} for a recent review). In this work, they gave the electron a form factor
\begin{equation}
\label{Eq:form}
\rho(\omega)= \frac{\Omega^2}{\omega^2+\Omega^2}
\end{equation}
with $\Omega$ a cut-off frequency. For the point electron, $\Omega\rightarrow \infty$, this recovers the Abraham-Lorentz equation, and decreasing the cut-off frequency produces related equations with the same pathologies. For the critical value $\Omega=\tau^{-1}$, however, the third order derivatives cancel, and Ford and O'Connell recovered (\ref{Eq:Eliezer-Ford-O'Connell-nonrelativistic}). It is worth noting that for still smaller cut-offs, third order derivatives reappear, yet the pathologies do not: despite many claims to the contrary, it is not the order of the Lorentz-Dirac equation that causes problems, but its precise form.

At first glance, (\ref{Eq:Eliezer-Ford-O'Connell}) appears to be no different from the Landau-Lifshitz equation. Indeed, their similarity has caused a certain confusion in the literature \cite{OConnell}. However, if $f_{\rm ext}^a$ represents the Lorentz force, its derivative introduces terms proportional to acceleration. Since the Landau-Lifshitz equation neglects terms of order ${\cal O}(\tau^2)$, these acceleration terms can be replaced by the Lorentz force without further loss of accuracy. However, since the Eliezer-Ford-O'Connell equation is regarded as exact, we are forced to retain these terms, and end up with a matrix equation for the acceleration:
\begin{equation}
\label{Eq:Eliezer-Ford-O'Connell-matrix}
\big(\delta^a_b+ \tau\qm \Delta^a{}_c F^c{}_b \big)\xdd^b=-\qm \big(F^a{}_b+\tau \xd^c\partial_c F^a{}_b\big)\xd^b
\end{equation}
where, for notational convenience, the label `ext' has been omitted from the external field $F^{\rm ext}_{ab}$. We will adopt this convention for the remainder of this chapter.

It can be shown \cite{Kravets} that (\ref{Eq:Eliezer-Ford-O'Connell-matrix}) can always be solved for $\xdd^a$, so the Eliezer-Ford-O'Connell equation is mathematically viable. Nevertheless, its validity remains open to question. Firstly, (\ref{Eq:Eliezer-Ford-O'Connell-nonrelativistic}) relies on the specific form factor (\ref{Eq:form}) with $\Omega=\tau^{-1}$ { (though Ford and O'Connell argue that it should be a good approximation for any form factor~\cite{FO1})}, while there is as yet no evidence that the electron is anything other than a point particle. Moreover, (\ref{Eq:Eliezer-Ford-O'Connell}) exists only as the relativistic generalization of (\ref{Eq:Eliezer-Ford-O'Connell-nonrelativistic}), and it is unclear whether a fully consistent derivation could exist.

Regardless of its validity as a description of radiating electrons, the Eliezer-Ford-O'Connell equation can serve a useful purpose. Since it { also appears} as an intermediate step in the derivation of the Landau-Lifshitz equation from the Lorentz-Dirac equation, it can be used to test the validity of the former: where the Landau-Lifshitz and Eliezer-Ford-O'Connell equations disagree, we cannot trust the Landau-Lifshitz equation.

\subsection{Mo--Papas equation}

In 1971, Mo and Papas proposed a new equation of motion \cite{MoPapas} for a radiating particle which they hoped would overcome the problems of the Lorentz-Dirac equation. Rather than working from first principles, they argued heuristically that such an equation should have certain features: it should depend only on the applied field and the particle's worldline; it should balance inertia and radiation forces with the Lorentz force and an additional acceleration-dependent generalization of the Lorentz force. This led them to postulate the equation
\begin{equation}
\label{Eq:MoPapas}
\xdd^a-\tau \qm F^b{}_c\xdd_b\xd^c\xd^a= -\qm F^a{}_b\xd^b+ gF^a{}_b\xdd^b.
\end{equation}
Here, the second term on the LHS compensates for the energy-momentum lost to radiation, while the second term on the RHS is their new force. To ensure consistency with the normalization condition, they took $g=-\tau q/m$.

While their motivation is clear, there is little in Mo and Papas's work that inexorably leads to (\ref{Eq:MoPapas}). Why approximate the Larmor power $m\tau \xdd_a \xdd^a$ by $\tau q F^a{}_b\xd^b\xdd_a$? Why introduce the mysterious acceleration analogue of the Lorentz force? And so it is remarkable that they should end up with an equation so close to Eliezer-Ford-O'Connell. Indeed, the only difference between the two equations is the term in the latter involving derivatives of the field.

The Mo--Papas equation quickly generated a certain level of interest. But it has been criticized on a number of fronts. One particular objection is that, for purely linear motion, it reduces to the ordinary Lorentz force, the additional terms precisely cancelling out. In general, radiation reaction is expected to be only a minor correction in the case of linear motion, but nevertheless it would be a surprise if it vanished identically.

\subsection{Bonnor equation}

Shortly after the work of Mo and Papas, a more radical proposition was put forward by Bonnor \cite{Bonnor}. Since the third order Schott term in the Lorentz-Dirac equation was both the source of the pathological behaviour and the least intuitive contribution, he suggested it should be dropped, yielding the equation of motion
\begin{equation}
\label{eq:Bonnor}
\frac{d (m\xd^a)}{d\lambda}=-qF^a{}_b\xd^b - \frac{q^2}{6\pi} \xdd_b \xdd^b\, \xd^a.
\end{equation}
Note that we have here not written the prefactor of the radiation reaction force as $m\tau$. This is because consistency with the normalization condition $\xd_a \xd^a =-1$ requires the particle's rest mass $m$---and consequently also $\tau$---to vary with time. Expanding the derivative in (\ref{eq:Bonnor}) and contracting with $\xd$ yields
\begin{align}
\label{eq:mass} \frac{dm}{d\lambda}&=-\frac{q^2}{6\pi} \xdd_a \xdd^a,\\
\label{eq:accel} m\xdd^a&=-q F^a{}_b\xd^b.
\end{align}

The interpretation of (\ref{eq:mass}-\ref{eq:accel}) is immediate: the energy lost to radiation is provided by a decrease in the particle's mass-energy, while the acceleration of the particle is governed by the usual Lorentz equation, albeit with a varying mass. It follows that in regions where the external field vanishes the acceleration is zero and the mass is constant, so the pathologies of the Lorentz-Dirac equation are again avoided. However, (\ref{eq:Bonnor}) introduces a new problem not faced by the Lorentz-Dirac equation.

Since acceleration is spacelike ($\xdd_a \xdd^a >0$) it follows from (\ref{eq:mass}) that the mass can only decrease, never increase. Thus the mass of a particle depends on its entire history. However, experiments show that all electrons have the same mass, to an extraordinarily high precision. In addition, comparing with (\ref{eq:accel}) shows that the smaller the mass becomes, the faster it decreases, for a given external force. Eventually the particle will radiate away all its mass, at which point it should travel at the speed of light. However, since by construction (\ref{eq:Bonnor}) preserves the normalization of $\xd^a$, this cannot be the case. For all its elegance, it seems the Bonnor equation cannot be a valid description of radiating particles.

\subsection{Sokolov equation}

More recently, Sokolov \cite{Sokolov} has introduced an equation that also departs radically from a conventional tenet of physics, in this case that the 4-momentum should be collinear with the 4-velocity. The justification for doing so is that part of the momentum of a charged particle may be regarded as distributed throughout space in its Coulomb field. Since this does not change instantaneously when the particle's motion is disturbed, it can be argued that an accelerating electron has a momentum and a velocity that are not parallel.

If we take momentum and velocity to be parallel, $p^a=m\xd^a$, the Einstein relation for energy and momentum is equivalent to parametrization by proper time:
\begin{equation}
\label{Eq:Einstein}
E^2-{\bm p}^2=m^2 \iff \xd^a\xd_a=-1.
\end{equation}
However, if we accept that acceleration causes $p^a$ and $\xd^a$ to be misaligned, in general only one of the equations in (\ref{Eq:Einstein}) can hold. Keeping the Einstein relation, we are led (via some quite general assumptions) to the equations
\begin{align}
\label{Eq:Sokolov-x}
\xd^a&= (\delta^a_b - \tau \qm F^a{}_b) \frac{p^b}{m},\\
\label{Eq:Sokolov-p}
\frac{\dot{p}^a}{m}&= -\qm F^a{}_b \frac{p^b}{m}+ \tau \frac{q^2}{m^2}\Big( F^a{}_bF^b{}_c\frac{p^c}{m}+ F^b{}_d F^d{}_c \frac{p_b}{m} \frac{p^c}{m}\frac{p^a}{m} \Big).
\end{align}

Apart from terms involving the derivatives of the fields, the form of (\ref{Eq:Sokolov-p}) is identical to the Landau-Lifshitz equation, under the substitution $p^a/m\rightarrow \xd^a$, though its derivation is quite different. The novel feature of the Sokolov theory is (\ref{Eq:Sokolov-x}), which describes the non-collinearity of $\xd^a$ and $p^a$.

Since its inception, the Sokolov theory has gained significant attention (though it is still far from universally accepted). However, it should be noted that this theory too suffers a number of drawbacks, stemming from abandoning the normalization condition in (\ref{Eq:Einstein}). While $E^2-{\bm p}^2=m^2$ has a clear physical meaning, parametrizing the worldline by proper time is simply a choice, and one that we are always free to make. The physical meaning of the parametrization used in (\ref{Eq:Sokolov-x}-\ref{Eq:Sokolov-p}) is obscure, and it is unclear why this should naturally emerge.

Using (\ref{Eq:Sokolov-x}), the normalization of velocities becomes
\begin{equation}
-\xd^a\xd_a= \Big(1-\tau^2 \frac{q^2}{m^2} F^a{}_bF_{ac} \frac{p^b}{m}\frac{p^c}{m}\Big) \leq 1.
\end{equation}
In general we could use this to rewrite (\ref{Eq:Sokolov-x}-\ref{Eq:Sokolov-p}) in terms of proper time $\lambda$. However, for sufficiently large fields and/or high energies, we could have $\xd^a\xd_a\geq 0$. Then the notion of proper time breaks down, and we find that a massive particle must move at the speed of light (or faster!). This is the complement to the problem faced by Bonnor's equation, and demonstrates that the Sokolov theory is capable in extreme circumstances of violating causality.
\section{Collective effects}
Although thus far we have focussed on the behaviour of a single particle in an externally applied field $F^{\rm ext}_{ab}$, in practice radiation reaction is unlikely to ever be observed in the context of a single radiating particle. However, modern laser facilities accelerate electron bunches with charge of the order of $10\,{\rm pC}$, containing $10^8$ particles, and modelling such a large number of particles requires an efficient mathematical description.

The most common approach to describing a large collection of slowly-moving point charges begins with the Liouville equation for an $N$-particle probability distribution. A cluster decomposition is used to express the $N$-particle probability distribution in terms of reduced $M$-particle distributions, where $M<N$, leading to a set of integro-differential equations for the reduced $M$-particle distributions (the {\it BBGKY hierarchy}). However, the full set of equations is intractable and physical reasoning must be used to cast the BBGKY hierarchy into a manageable form. If all correlations between particles are negligible then the Vlasov equation
\begin{equation}
\label{Vlasov}
\partial_t f + {\bm \nabla}_{\bm x}\cdot ({\bm v} f) + {\bm \nabla}_{\bm v}\cdot \bigg[\frac{1}{m}{\bm F}\,f\bigg] = 0
\end{equation}
is obtained for the $1$-particle distribution $f(t,{\bm x},{\bm v})$ where the electrostatic force ${\bm F} = - q{\bm \nabla}V$. The mean electric potential $V$ is determined by Poisson's equation
\begin{equation}
\label{Poisson}
{\bm\nabla}^2 V = - \rho/\varepsilon_0
\end{equation}
where the electric charge density $\rho$ is
\begin{equation}
\label{charge_density}
\rho(t,{\bm x}) = q \int f(t,{\bm x},{\bm v}) d^3 v.
\end{equation}
An introduction to the BBGKY hierarchy may be found in Ref.~\cite{boyd:2003}.
 
Alternatively, from a purely mathematical perspective, one can rigorously show that the behaviour of a collection of $N$ point charges interacting via their Coulomb fields is described by the Vlasov-Poisson system of equations in the limit $N\rightarrow\infty$ (see Ref.~\cite{kiessling:2008} for a recent review).

The non-relativistic Vlasov-Maxwell system is obtained by the replacement ${\bm F} = - q {\bm \nabla}V \rightarrow q({\bm E} + {\bm v} \times {\bm B}$) in (\ref{Vlasov}) where the mean electric field ${\bm E}$ and mean magnetic field ${\bm B}$ satisfy Maxwell's equations
\begin{align}
&{\bm \nabla}\cdot{\bm E} = \rho/\varepsilon_0,\qquad {\bm \nabla}\times{\bm E} = - \partial_t {\bm B},\\
&{\bm \nabla}\times{\bm B} = \frac{1}{c^2} \partial_t {\bm E} + \mu_0 {\bm J},\qquad {\bm \nabla}\cdot{\bm B} = 0
\end{align}
and the electric current density ${\bm J}$ is
\begin{equation}
\label{electric_current}
{\bm J}(t,{\bm x}) = q \int {\bm v}f(t,{\bm x},{\bm v}) d^3 v.
\end{equation}
Although it is straightfoward to motivate the {\it relativistic} Vlasov-Maxwell system by rendering (\ref{Vlasov}, \ref{charge_density}, \ref{electric_current}) invariant under Lorentz transformations, we are unaware of any mathematically rigorous derivation of the relativistic Vlasov-Maxwell system that begins with a set of point particles interacting via their Li\'enard-Wiechert fields.

Use of the Vlasov-Maxwell and Vlasov-Poisson systems is ubiquitous in plasma physics and particle accelerator physics. Although the relativistic Vlasov-Maxwell system is often given the appellation ``self-consistent'', this should not be misconstrued to mean that it incorporates the recoil that each particle experiences due to its own emission. At present, there is no universally accepted (physical or mathematical) approach to a tractable kinetic theory starting from first principles that describes a bunch of relativistic particles each of which is reacting to its own emission as well as the electromagnetic fields of the other particles.

A pragmatic approach introduced in the context of magnetized plasmas~\cite{berezhiani:2004} and tokamaks~\cite{hazeltine:2004}, and recently adopted by the laser-plasma community~\cite{tamburini:2011}, is to simply augment the Lorentz force in the Vlasov equation with the Landau-Lifshitz radiation reaction force due to the mean electromagnetic field and an externally applied electromagnetic field. Insight into this approach may be obtained by noting that the generalized Vlasov equation can be understood as the preservation of the product of the $1$-particle distribution and a ``volume'' element (a differential form of maximal degree) along the orbits of the Landau-Lifshitz equation on the single-particle ``phase'' space coordinated by $(t,{\bm x},{\bm v})$. By including an acceleration coordinate ${\bm a}$, one can also follow a similar prescription using the Lorentz-Dirac equation~\cite{noble:2013} and exploit advantages in delaying the removal of runaway behaviour and pre-acceleration to later in the analysis. Following the approach of the latter case, consideration of the Lorentz-Dirac equation yields
\begin{equation}
\label{generalized_vlasov}
Lf + \frac{3}{\tau} f = 0 
\end{equation}
for the $1$-particle distribution $f= f(x,\bm{v},\bm{a})$ where $L$ is the Liouville operator
\begin{align}
L = \dot{x}^a \frac{\partial}{\partial x^a} + a^\mu \frac{\partial}{\partial v^\mu} + \bigg[\ddot{x}^a \ddot{x}_a v^\mu + \frac{1}{\tau}\bigg (a^\mu + \frac{q}{m} F^\mu{ }_a \dot{x}^a\bigg) \bigg]\frac{\partial}{\partial a^\mu},
\end{align} 
$(x)$ is shorthand for $(t,{\bm x})$ and Greek indices $\mu, \nu$ range over $1, 2, 3$ and are raised and lowered using the Kronecker delta $\delta^\mu_\nu$.  The $4$-velocity coordinate $\dot{x}^a$ satisfies $ \dot{x}^a \dot{x}_a = -1$ and is given in terms of the proper velocity ${\bm v}$ as $\dot{x}^0 = \sqrt{1+{\bm v}^2}$, $\dot{x}^\mu = v^\mu$ where $\bm{v}^2 = v^\mu v_\mu$. Likewise, the $4$-acceleration coordinate $\ddot{x}^a$ satisfies $\ddot{x}^a \dot{x}_a = 0$ and is given in terms of ${\bm a}$ and ${\bm v}$ as $\ddot{x}^0 = {\bm a} \cdot {\bm v}/\sqrt{1+{\bm v}^2}$ , $\ddot{x}^\mu = a^\mu$.
 
The second term on the left-hand side of (\ref{generalized_vlasov}) may be understood as a consequence of losses due to radiation. 

Maxwell's equations for the mean field $F_{ab}$ are
\begin{align}
\label{cov_faraday}
&\partial_a F_{bc} + \partial_c F_{ab} + \partial_b F_{ca} = 0,\\
\label{cov_amp-max}
&\partial_a F^{ab} = J^b + J^b_{\rm ext}
\end{align}
with $J^a$ the electric $4$-current
\begin{equation}
\label{number_current}
J^a(x) = q \int \dot{x}^a f(x,{\bm v},{\bm a})\, \frac{d^3 v\, d^3 a}{1 + \bm{v}^2}.
\end{equation}
The presence of the factor $1+\bm{v}^2$ in (\ref{number_current}) and the second term in (\ref{generalized_vlasov}) are related, as discussed in Ref.~\cite{noble:2013}, and $J^a_{\rm ext}$ is an external $4$-current. 

Almost all solutions to (\ref{generalized_vlasov}) will exhibit the pathological behaviour inherent in the Lorentz-Dirac equation as described earlier. Physically acceptable solutions may be extracted from (\ref{generalized_vlasov}) using the ansatz
\begin{equation}
\label{physical_constraint}
f(x,{\bm v},{\bm a}) = \sqrt{1+{\bm v}^2} g(x,{\bm v})\, \delta^{(3)}\big(\bm{a} - \bm{A}(x,\bm{v})\big)
\end{equation}
where $\delta^{(3)}$ is the $3$-dimensional Dirac delta and $g(x,{\bm v})$, ${\bm A}(x,\bm{v})$ are assumed to have a power-series dependence on $\tau$: 
\begin{align}
\label{g_tau_ansatz}
&g(x,{\bm v}) = \sum\limits^{\infty}_{n=0} g_{(n)}(x, {\bm v})\,\tau^n,\\
\label{A_tau_ansatz}
&\bm{A}(x,\bm{v}) = \sum\limits^{\infty}_{n=0} \bm{A}_{(n)}(x, {\bm v})\,\tau^n.
\end{align}
The subspace $(x, {\bm v}) \mapsto (x, {\bm v}, {\bm a} = {\bm A}(x, {\bm v}))$ of $(x,{\bm v},{\bm a})$ space contains physical solutions to the Lorentz-Dirac equation and the factor $\sqrt{1+\bm{v}^2}$ ensures that $g$ is normalized in the usual manner for a $1$-particle distribution; plugging (\ref{physical_constraint}) into (\ref{number_current}) yields the usual expression for the electric $4$-current in relativistic kinetic theory:
\begin{equation}
J^a(x) = q\int \dot{x}^a g(x,\bm{v})\, \frac{d^3 v}{\sqrt{1 + \bm{v}^2}}.
\end{equation}
Equations (\ref{physical_constraint}, \ref{generalized_vlasov}) lead to the coupled system of equations
\begin{align}
\label{physical_kinetic_1}
&\dot{x}^a\frac{\partial A^\mu}{\partial x^a} +  A^\nu \frac{\partial A^\mu}{\partial v^\nu} = A^a\,A_a\, v^\mu + \frac{1}{\tau}(A^\mu + \frac{q}{m}F^\mu{ }_a \dot{x}^a),\\
\label{physical_kinetic_2}
&\dot{x}^a \frac{\partial g}{\partial x^a} + \sqrt{1+\bm{v}^2}\frac{\partial}{\partial v^\mu}\bigg(\frac{g\,A^\mu}{\sqrt{1+\bm{v}^2}}\bigg) = 0
\end{align}
for $g$ and $\bm{A}$, with $A^0 = v^\mu A_\mu / \sqrt{1+\bm{v}^2}$.

Analysis of (\ref{physical_kinetic_1}, \ref{physical_kinetic_2}) shows that neglecting ${\cal O}(\tau)$ terms in (\ref{g_tau_ansatz}, \ref{A_tau_ansatz}) leads to the usual relativistic Vlasov equation without the self-force. Neglecting ${\cal O}(\tau^2)$ leads to the kinetic theory of the Landau-Lifshitz equation as found in, for example, Ref. \cite{tamburini:2011}. Furthermore, it may be shown~\cite{noble:2013} that the entropy $4$-current $s^a$ defined as
\begin{equation}
\label{kinetic_entropy_def}
s^a = - k_B \int \dot{x}^a\,g \ln(g)\, \frac{d^3 v}{\sqrt{1+\bm{v}^2}}
\end{equation}
satisfies
\begin{align}
\label{entropy_div}
\partial_a s^a =& -\tau \frac{k_B}{m} \bigg(J_a (J^a + J^a_{\text{ext}}) + 4 \frac{q^2}{m^2}T_{ab} S^{ab}\bigg) + {\cal O}(\tau^2)
\end{align}
where
\begin{align}
\label{stress_definition}
&S^{ab} = m\int \dot{x}^a \dot{x}^b g\frac{d^3 v}{\sqrt{1+\bm{v}^2}},\\
&T^{ab} = F^{ac} F^b{ }_c - \frac{1}{4}\eta^{ab} F_{cd} F^{cd}
\end{align}
with $S^{ab}$ the stress-energy-momentum tensor of the matter encoded by $g$ and $T^{ab}$ is the stress-energy-momentum tensor of the electromagnetic field $F_{ab}$.

The right-hand side of (\ref{entropy_div}) naturally splits into two contributions of opposite sign when $J^a_{\text{ext}}=0$. The first term $-J_a J^a$ is positive due to the signature choice $\eta_{ab} = \text{diag}(-1,1,1,1)$ and encourages the entropy to increase. However, $T_{ab}$ satisfies the weak energy condition $T_{ab} X^a X^b \ge 0$ for all choices of timelike vector $X^a$, so $T_{ab} \dot{x}^a \dot{x}^b \ge 0$ and $-T_{ab} S^{ab} \le 0$ follows from (\ref{stress_definition}) since the $1$-particle distribution $g$ is positive. The first term on the right-hand side of (\ref{entropy_div}) drives growth (heating) in the phase-space volume occupied by the system of particles, whereas the second term drives a reduction (cooling) of the system's phase-space volume.
 
If the self-fields of the particles are neglible relative to the applied external field (as is the case for a bunch of electrons driven by an ultra-intense laser pulse) then $J_a J^a + 4 q^2 T_{ab} S^{ab}/m^2$ may be replaced by $4 q^2 T^{\text{ext}}_{ab} S^{ab}/m^2$ where $T^{\text{ext}}_{ab}$ is the stress-energy-momentum tensor of the externally applied field $F^{\text{ext}}_{ab}$. Thus, a sufficiently strong field $F^{\text{ext}}_{ab}$ will encourage a charged particle beam to cool. However, the situation is considerably more subtle when the self-fields dominate because it is then possible for the beam to heat due to radiation reaction. A discussion of the implications of (\ref{entropy_div}) for an {\it isolated} bunch of electrons is provided in Ref.~\cite{burton:2013}. 

Most attention has been focussed in recent years on the behaviour of a bunch of particles driven by one or more ultra-intense laser beams. In such situations the discrete nature of charge can induce growth in the phase space volume (so-called {\it stochastic heating}). However, the interplay of the dissipation due to the radiation reaction force and stochastic heating~\cite{lehmann:2012} can lead to an improvement in the monochromaticity of particle bunches accelerated by multiple ultra-intense laser beams.
\section{Quantum considerations}
It is often remarked that, being so brief, the time $\tau$ belongs to the quantum realm, and so we should not worry if the classical theory predicts violations of causality over this timescale. Although rather vague---$\tau$ itself is a purely classical constant, being independent of $\hbar$---some credence can be given to this by comparing it to the Compton wavelength $\lambda_c=h/mc$:
\begin{equation}
c\tau= \frac{q^2}{6\pi \varepsilon_0 m c^2}= \frac{2}{3}\frac{\alpha}{2\pi} \lambda_c,
\end{equation}
where $\alpha=q^2/4\pi\varepsilon_0 \hbar c\approx 1/137$ is the fine structure constant. The distance light can travel in a time $\tau$ is less than one part in a thousand of the Compton wavelength, a lengthscale regarded as firmly within the quantum realm. Thus, it is reasonable to ask if a quantum mechanical treatment can elucidate the issues surrounding the Lorentz-Dirac equation.

\subsection{Eliminating pathologies}
One of the earliest explorations of radiation reaction in a quantum context was given by Moniz and Sharp \cite{Moniz}. Noting that in the classical theory an extended charge is not subject to runaways or pre-acceleration provided its radius exceeds $c\tau$ \cite{Levine}, they asked whether the Compton wavelength in the quantum theory might in the same way ameliorate the pathological behaviour. Put another way, could the quantum uncertainty in the particle's position give rise to an effective radius $\lambda_c\gg c\tau$?

Analogously to Lorentz's derivation of the equation of motion for an extended classical charge, Moniz and Sharp derived the Heisenberg equation of motion for the position operator of an extended electron:
\begin{equation}
\label{eq:Moniz}
m\frac{d^2 {\bm R}}{dt^2}={\bm F}_{\rm ext}+ \sum^\infty_{n=2} A_n \frac{d^n {\bm R}}{dt^n},
\end{equation}
where
$A_n$ are constants depending on the Compton wavelength and the particle's charge density.

To get from (\ref{eq:Moniz}) to an equation for a classical point electron, two limits must be taken. If we first take the classical limit $\hbar\rightarrow 0$ (here equivalent to $\lambda_c\rightarrow 0$), we recover Lorentz's theory, which reduces to the Abraham-Lorentz equation when the point particle limit is taken. However, Moniz and Sharp found that, if they first took the point particle limit and only then the classical limit, the situation was quite different. Although they were not able to obtain an equation of motion in closed form, they were able to show that the theory was free from runaways and pre-acceleration.

Unfortunately, major problems arise in generalising this result to the relativistic domain. Firstly, the calculations are far more cumbersome when relativistic and quantum effects are simultaneously present, and such nonlinear processes as pair production come into play. Moreover, in their demonstration that pathologies are absent, Moniz and Sharp had to appeal to consistency with the assumption of nonrelativistic behaviour. It is far from clear, therefore, whether the same results are true in the relativistic theory, where radiation reaction is particularly important.

\subsection{Classical limit of QED}
Although there were earlier investigations of radiation reaction in relativistic quantum electrodynamics, Higuchi and Martin \cite{Higuchi1,Higuchi2,Higuchi3} conducted the first extensive comparison between the predictions of classical electrodynamics and the classical limit of relativistic QED. By analysing the expectation value of the position of an electron wave-packet after interacting with an external potential, they found that in the classical limit this agreed with the result obtained from the Landau-Lifshitz equation. Since they worked to leading order in the coupling $\alpha$ (i.e., considered single photon emission only), this is consistent with the Lorentz-Dirac equation.

More recently, Ilderton and Torgrimsson \cite{Ilderton1,Ilderton2} have similarly investigated the classical limit of QED, this time in the physically relevant background of a plane electromagnetic wave. In contrast to Higuchi and Martin, they considered the expectation value of the 4-momentum operator, rather than the position shift, and traced its value throughout the interaction, not only in the asymptotic limit. Again, they found agreement with the Landau-Lifshitz equation, though they also noted that, to this order, their results were consistent with the Eliezer-Ford-O'Connell equation, but not with the theories of Sokolov or Mo and Papas.

Higuchi and Martin worked to tree level, but pointed out that 1-loop effects could be relevant to the classical limit. Ilderton and Torgrimsson showed this explicitly, finding that Feynman diagrams corresponding to emission and absorption of a single photon were necessary to cancel divergences in processes involving radiation of a photon that is not subsequently reabsorbed. This demonstrates the intimate connection between `radiation reaction' and `self-force', terms that are often used synonymously.

\subsection{Quantum effects}
So far, we have addressed the classical limit of various quantum approaches to radiation reaction. However, it is important to note that quantum effects can be important in their own right. In many cases, radiation reaction can act to prevent quantum modifications to particle motion from becoming significant: in general, the classical theory is sufficient unless both high energies and strong fields are present, and radiation losses can ensure that an electron's energy is (relatively) low when it accesses a region of high field. However, this is not always the case, and we now examine some of the ways in which quantum radiation reaction can be distinct from the classical effects.

{ QED is formulated by promoting the classical electromagnetic field and the Dirac spinor field to quantum operators. In strong-field QED, the classical electromagnetic field is split into two parts prior to quantization : a background term (such as a strong laser pulse), which remains a classical field, and a perturbation to the background field. Only the perturbation is promoted to a quantum operator, so the perturbation alone is the photon field. The electron-positron quantum states are constructed using a basis of exact solutions to the Dirac equation in the background field and perturbation theory is used to calculate matrix elements between quantum states in the background field. However, only a handful of electromagnetic fields exist for which this can be achieved (typical examples of background fields include a static magnetic field, or a plane electromagnetic wave) and applying this formalism to a realistic laser pulse remains a formidable challenge.

Considerable interest in QED in strong magnetic fields grew during the 1960s as a consequence of the multi-${\rm MG}$ field strengths offered by explosive flux compression techniques developed at that time. Many of the results obtained during that era~\cite{erber:1966} have been used in contemporary theoretical studies of quantum radiation reaction in ultra-intense laser-matter interactions, such as the work by Bell, Kirk, et al.~\cite{bell:2008, blackburn:2014}. From an heuristic perspective, the practical features of quantum physics are that it is discrete and random, in contrast to the continuous and deterministic nature of the classical world, and it is these features that are responsible for many of the differences between the quantum and classical predictions of radiation reaction. 
}

Quantum effects must be accounted for when the ratio of the electric field `seen' by the electron is comparable to the Sauter-Schwinger field, $E_S= m^2c^3/|q|\hbar$:
\begin{equation}
\chi:= \frac{E_e}{E_S}= \frac{|q|\hbar}{m^2}\sqrt{F_{ab}F^{ac}\xd^b\xd_c} \gtrsim 1.
\end{equation}
Heuristically, the electron proceeds without losing energy in such fields except at discrete emission events, where its motion is altered significantly. This means it can penetrate deeply into a laser pulse, say, before it emits any radiation, whereas the classical picture would have it radiating---and hence losing energy---as soon as it enters the pulse. As such, the classical theory tends to overestimate the significance of radiation reaction, compared to the (more accurate) quantum description. Di Piazza {\it et al.} have demonstrated this explicitly in the case of multi-photon Compton scattering \cite{DiPiazza}.

In some cases, quantum effects can not only reduce the classical radiation reaction effects, but reverse them completely. For example, it is clear from the Lorentz-Dirac equation that more energetic particles tend to radiate more than less energetic ones. In the absence of quantum effects this leads to a reduction in the energy spread of an electron bunch (assuming repulsive interparticle forces are negligible compared to the radiation reaction force, which is reasonable for an ultrarelativistic bunch). Similar observations are true of the spread in momentum. As quantum effects become more significant, however, the continuous radiation driving solutions to the Lorentz-Dirac equation gives way to discrete, stochastic emission events, and this stochasticity tends to increase the spread in energy-momentum. In laser-particle interactions where quantum effects are important but pair production remains negligible, it has recently been shown that this can more than compensate for the classical reduction in both longitudinal \cite{Neitz} and transverse \cite{Green} momentum spread, leading to degradation in the quality of electron bunches. For further information on the quantum description of intense laser-particle interactions, we refer the reader to the recent review article \cite{DiPiazza_Review}.

{
We close this section with a brief comment on the intriguing interplay between quantum fluctuations, dissipation and radiation reaction.

One of the first major triumphs of QED was the calculation of the Lamb shift in the energy of the ${}^2S_{1/2}$ orbital of the hydrogen atom. Early explanations of this effect \cite{Welton} were based on the interaction of the atom with vacuum fluctuations in the electromagnetic field. However, from the fluctuation-dissipation theorem, it is clear that such fluctuations can be related to dissipative processes, which Ford, Lewis and O'Connell \cite{FLO1, FLO2} have shown can describe radiation reaction.

In the 1970s, Ackerhalt, Knight and Eberly \cite{AKE} presented a novel calculation of the Lamb shift in which the entire shift, and the concomitant spectral line broadening, resulted from radiation reaction, with the contribution from vacuum fluctuations cancelling out. The discrepancy between this and earlier interpretations was explained by Milonni, Ackerhalt and Smith \cite{MAS} as resulting from alternative orderings of the atomic and field operators.  These operators commute, so the physical predictions are unaffected, but their partition into free and interacting operators does not reflect this commutativity. As a result, an interaction between the vacuum field and the excitation of the atom (vacuum fluctuation effect) can be converted into an interaction between the atom and the radiation it produces (radiation reaction effect) purely by a reordering of commuting operators. Thus, one must be careful when attributing a given effect to radiation reaction (or any specific physical process), since the physical cause is not always uniquely defined.
} 
\section{Experimental signatures}
As noted in the Introduction, the question of radiation reaction has vexed theorists for over a century, but has yet to trouble their experimentalist colleagues. It may be that the lack of experimental support is the chief reason why a satisfactory theoretical understanding of the problem has yet to emerge. There is historical precedent for this: for example, the mathematical structures behind the renormalization programme in quantum electrodynamics had been in place for a decade before Lamb and Retherford's measurement of the fine structure of hydrogen provided the stimulus for Schwinger, Feynman and Tomonaga to complete the picture.

With the recent advances in laser technology, experimental investigation of radiation reaction phenomena is at last becoming a realistic prospect. For example, ELI, due to come online in 2017, is expected to operate at intensities of \mbox{$10^{23}$ W cm$^{-2}$}, with GeV electrons \cite{ELI}. Under these conditions, the radiation reaction force is comparable to, and can even exceed, the Lorentz force due to the laser. 

The first experimental signature of radiation reaction was derived by Dirac, in his original paper on the subject \cite{dirac:1938}. He found that, for an electron interacting with light of frequency $\omega$, the Thomson scattering cross-section, $\sigma_T=8\pi r^2_e/3$, where $r_e\simeq 2.8$ fm is the classical electron radius, is reduced by radiation reaction:
\begin{equation}
\label{eq:Thomson}
\sigma^\prime_T= \frac{\sigma_T}{1+\tau^2\omega^2}.
\end{equation}
Although the dependence of the cross-section on the frequency of the light makes it an appealing candidate for detecting radiation reaction, for optical lasers the deviation from $\sigma_T$ is only about 1 part in $10^{16}$. Additional complications arise from competing effects due to nonlinearities in the laser intensity, and quantum effects \cite{Heinzl:Thomson}. Nevertheless, shining laser pulses onto electrons has become the foremost set-up to measure radiation reaction.

\begin{figure}
\begin{center}
\includegraphics[height=5cm]{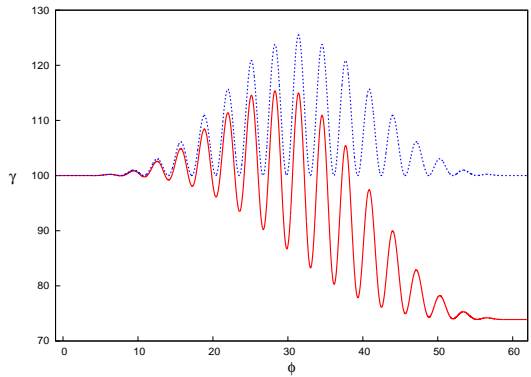}
\end{center}
\caption[example] 
{\label{fig:energy} 
\footnotesize Energy loss for an electron of initial Lorentz factor $\gamma=100$ interacting with a 10 cycle laser pulse of strength $a_0=100$. Blue dotted curve without radiation reaction; red solid curve with radiation reaction.}
\end{figure}

In recent years there has been an intensification of activity in the analysis of interactions of electrons and intense laser pulses \cite{DiPiazza:Landau-Lifshitz,Hadad,Lehmann:gain,Harvey:symmetry,Kravets}. In the absence of radiation reaction, under certain quite general conditions, an electron leaves a laser pulse with the same energy and momentum with which it entered. However, radiation reaction breaks the symmetry of the interaction. In the ultrarelativistic limit, the radiation reaction force is dominated by the last term in the Lorentz-Dirac equation, which can be written
\begin{equation}
f^a = -m\tau \xdd_b \xdd^b\, \xd^a.
\end{equation}
This clearly acts as a frictional force, whose strength goes like the square of the proper acceleration. Thus we expect a relativistic electron to exit a pulse with less energy and longitudinal momentum than it had initially. A typical example of this is shown in Fig.~\ref{fig:energy}.

It is clear from the Lorentz-Dirac equation that the radiation reaction force scales with the frequency $\omega$ of the laser, but since a relativistic particle sees a Doppler shifted laser field, this is enhanced by its Lorentz factor $\gamma$. Additionally, since the Landau-Lifshitz equation is quadratic in the fields, these effects also scale with the square of the intensity parameter, $a_0= {q\cal E}/m\omega c$. Thus a measure ${\cal R}$ of the significance of radiation reaction is given by
\begin{equation}
\label{eq:rrparameter}
{\cal R} = 2\tau\omega\gamma a^2_0.
\end{equation}
Do not be misled by the inverse factors of $\omega$ appearing in the definition of $a_0$: it is $a_0$ rather than $\cal E$ that is Lorentz invariant \cite{Heinzl:a0}, so {$\cal R$} increases linearly with~$\omega$. 

In general, radiation reaction effects on the radiation itself are of second order in $\tau$, and so are usually less pronounced than the effects on the electrons.
However, this does not mean that radiation reaction effects cannot be observed in the radiation spectrum. Since the radiation emitted by a relativistic particle is highly concentrated in the direction of its motion, small changes in this direction can radically alter the angular radiation distribution. It was noted in Ref.~\cite{DiPiazza:sub} that, provided
\begin{equation}
{\cal R} \gtrsim \frac{4\gamma^2-a^2_0}{2a^2_0},
\end{equation}
even if ${\cal R}\ll 1$ an electron can reverse its direction of motion in a laser pulse, an effect which does not occur in the absence of radiation reaction. This can greatly broaden the angular distribution of the radiation it emits, which may be the most accessible signature of radiation reaction.

{ Finally, it was recently argued~\cite{blackburn:2014} that an experiment using {\it current} high-energy laser facilities could be realized in which the consequences of the discrete nature of photon emission would be manifest. Monte Carlo simulations incorporating quantum radiation reaction show that the parameters of a collision between an electron beam and a high-energy laser pulse can be chosen for which the photon yield is considerably greater than the photon yield predicted by classical radiation reaction.}
\section{Conclusion}
Numerous theoretical proposals for describing the behaviour of matter in ultra-intense electromagnetic fields have been explored over a number of decades. However, it is clear that theory and experiment are now approaching a significant juncture. We will soon enter an era when it will be possible to experimentally investigate the behaviour of matter bombarded by ultra-intense lasers in regimes where the magnitudes of the radiation reaction force and Lorentz force are comparable. 
Such experiments will cast new light on fundamental questions concerning the behaviour of light and matter.
\section*{Acknowledgements}
This work was undertaken as part of the ALPHA-X consortium funded under EPSRC grant EP/J018171/1. DAB is also supported by the Cockcroft Institute of Accelerator Science and Technology  (STFC grant ST/G008248/1).


\begin{thebibliography}{99}
%
\bibitem{ELI} http://www.extreme-light-infrastructure.eu
%
\bibitem{kaplan:2002}
A.E Kaplan and P. L. Shkolnikov, {\it Lasetron: A Proposed Source of Powerful Nuclear-Time-Scale Electromagnetic Bursts}, Phys. Rev. Lett. 88 (2002), 074801.
%
\bibitem{erber:1961}
T. Erber, {\itshape The Classical Theories of Radiation Reaction}, Fortschr. Phys. 9 (1961), 343--392.
%
\bibitem{rohrlich:2007}
F. Rohrlich, {\itshape Classical Charged Particles}, 3rd Ed., World Scientific, 2007.
%
\bibitem{poisson:2011}
E. Poisson, A. Pound and I. Vega, {\itshape The Motion of Point Particles in Curved Spacetime}, Living Rev. Relativity 14 (2011), 7.
%
\bibitem{gralla:2009}
S.E. Gralla, A.I. Harte and R.M. Wald, {\itshape Rigorous derivation of electromagnetic self-force}, Phys. Rev. D 80 (2009), 024031.
%
\bibitem{teitelboim:1971}
C. Teitelboim, {\itshape Radiation Reaction as a Retarded Self-Interaction}, Phys. Rev. D 4 (1971), pp. 345--347.
%
\bibitem{barut:1974}
A.O. Barut, {\itshape Electrodynamics in terms of retarded fields}, Phys. Rev. D  10 (1974), pp. 3335--3336.
%
\bibitem{jackson:1999}
J.D. Jackson, {\itshape Classical Electrodynamics}, 3rd Ed., Wiley, 1999.
%
\bibitem{hammond:2010a}
R.T. Hammond, {\itshape Radiation reaction at ultrahigh intensities}, Phys. Rev. A 81 (2010), 062104.
%
\bibitem{hammond:2010b}
R.T. Hammond, {\itshape Relativistic Particle Motion and Radiation Reaction in Electrodynamics}, EJTP 7, No. 23 (2010), pp. 221--258.
%
\bibitem{landau:1987}
L.D. Landau and E.M. Lifshitz, {\itshape The Classical Theory of Fields (Course of Theoretical Physics, vol. 2)}, 4th Ed., Butterworth-Heinemann Ltd, 1987.
%
\bibitem{dirac:1938}
P.A.M. Dirac, {\itshape Classical theory of radiating electrons}, Proc. R. Soc. Lond. A 167 (1938), pp. 148--168.
%
\bibitem{ferris:2011}
M.R. Ferris and J. Gratus, {\itshape The origin of the Schott term in the electromagnetic self force of a classical point charge}, J. Math. Phys. 52 (2011), 092902.
%
\bibitem{Eliezer}
C.J. Eliezer, {\itshape On the classical theory of particles}, Proc. R. Soc. Lond. A 194 (1948), pp. 543--555.
%
\bibitem{FO1}
G.W. Ford and R.F. O'Connell, {\itshape Radiation reaction in electrodynamics and the elimination of runaway solutions}, Phys. Lett. A 157, (1991), 217--220.
%
\bibitem{FO2}
G.W. Ford and R.F. O'Connell, {\itshape Relativistic form of radiation reaction}, Phys. Lett. A 174 (1993), pp. 182--184.
%
\bibitem{oconnell:2012}
%
R.F. O'Connell, {\itshape Radiation reaction: general approach and applications, especially to electrodynamics}, Contemp. Phys. 53, 4, (2012), 301--313.
%
\bibitem{OConnell}
R.F. O'Connell, {\itshape The equation of motion of an electron}, Phys. Lett. A 313 (2003), pp. 491--497.
%
\bibitem{Kravets}
Y. Kravets, A. Noble and D.A. Jaroszynski, {\itshape Radiation reaction effects on the interaction of an electron with an intense laser pulse}, Phys. Rev. E 88 (2013), 011201(R).
%
\bibitem{MoPapas}
T.C. Mo and C.H. Papas, {\itshape New Equation of Motion for Classical Charged Particles}, Phys. Rev. D 4 (1971), pp. 3566--3571.
%
\bibitem{Bonnor}
W.B. Bonnor, {\itshape A new equation of motion for a radiating charged particle}, Proc. R. Soc. Lond. A 337 (1974), pp. 591--598.
%
\bibitem{Sokolov}
I.V. Sokolov, {\itshape Renormalization of the Lorentz–-Abraham–-Dirac equation for radiation reaction force in classical electrodynamics}, J. Exp. Theor. Phys. 16 (2009), pp. 207--212.
%
\bibitem{boyd:2003}
T.J.M. Boyd and J.J. Sanderson, {\it The Physics of Plasmas}, Cambridge University Press, 2003.
%
\bibitem{kiessling:2008}
M.K.H. Kiessling, {\itshape Microscopic derivations of Vlasov equations}, Commun. Nonlinear Sci. Numer. Simul. 13 (2008), pp. 106--113
%
\bibitem{berezhiani:2004}
V.I. Berezhiani, R.D. Hazeltine and S.M. Mahajan, {\itshape Radiation reaction and relativistic hydrodynamics}, Phys. Rev. E 69 (2004), 056406.
%
\bibitem{hazeltine:2004}
R.D. Hazeltine and S.M. Mahajan, {\itshape Radiation reaction in fusion plasmas}, Phys. Rev. E 70 (2004), 046407.
%
\bibitem{tamburini:2011}
M. Tamburini, F. Pegoraro, A. Di Piazza, C.H. Keitel, T.V. Liseykina and A. Macchi, {\itshape Radiation reaction effects on electron nonlinear dynamics and ion acceleration in laser-solid interaction}, Nucl. Instrum. Methods A 653 (1) (2011), pp. 181--185. 
%
%
\bibitem{noble:2013}
A. Noble, D.A. Burton, J. Gratus and D.A. Jaroszynski, {\itshape A kinetic model of radiating electrons}, J. Math. Phys. 54 (2013), 043101.
%
\bibitem{burton:2013}
D.A. Burton and A. Noble, {\itshape On the entropy of radiation reaction}, arXiv:1310.5906 (2013).
%
\bibitem{lehmann:2012}
G. Lehmann and K.H. Spatschek, {\itshape Phase-space contraction and attractors for ultrarelativistic electrons}, Phys. Rev. E 85 (2012), 056412.
%
\bibitem{Moniz}
E.J. Moniz and D.H. Sharp, {\itshape Absence of runaways and divergent self-mass in nonrelativistic quantum electrodynamics}, Phys. Rev. D 10 (1974), pp. 1133--1136.
%
\bibitem{Levine}
H. Levine, E.J. Moniz and D.H. Sharp, {\itshape Motion of extended charges in classical electrodynamics}, Am. J. Phys. 45 (1977), pp. 75--78.
%
\bibitem{Higuchi1}
A. Higuchi and G.D.R. Martin, {\itshape Lorentz-Dirac force from QED for linear acceleration}, Phys. Rev. D 70 (2004), 081701(R).
%
\bibitem{Higuchi2}
A. Higuchi and G.D.R. Martin, {\itshape Classical and quantum radiation reaction for linear acceleration}, Found. Phys. 35 (2005), pp. 1149--1179.
%
\bibitem{Higuchi3}
A. Higuchi and G.D.R. Martin, {\itshape Radiation reaction on charged particles in three-dimensional motion in classical and quantum electrodynamics}, Phys. Rev. D 73 (2006), 025019.
%
\bibitem{Ilderton1}
A. Ilderton and G. Torgrimsson, {\itshape Radiation reaction from QED: Lightfront perturbation theory in a plane wave background}, Phys. Rev. D 88 (2013), 025021.
%
\bibitem{Ilderton2}
A. Ilderton and G. Torgrimsson, {\itshape Radiation reaction in strong field QED}, Phys. Lett. B 725 (2013), pp. 481-486.
%
\bibitem{erber:1966}
T. Erber, {\itshape High-Energy Electromagnetic Conversion Processes in Intense Magnetic Fields}, Rev. Mod. Phys. 38 (1966), 626--659.
%
\bibitem{bell:2008}
A.R. Bell, J.G. Kirk, {\itshape Possibility of Prolific Pair Production with High-Power Lasers}, Phys. Rev. Lett. 101 (2008), 200403.
%
\bibitem{blackburn:2014}
T.G. Blackburn, C.P. Ridgers, J.G. Kirk and A.R. Bell, {\em Quantum Radiation Reaction in Laser-Electron-Beam Collisions}, Phys. Rev. Lett. 112 (2014), 015001. 
%
\bibitem{DiPiazza}
A. Di Piazza, K.Z. Hatsagortsyan and C.H. Keitel, {\itshape Quantum Radiation Reaction Effects in Multiphoton Compton Scattering}, Phys. Rev. Lett. 105 (2010), 220403.
%
\bibitem{Neitz}
N. Neitz and A. Di Piazza, {\itshape Stochasticity Effects in Quantum Radiation Reaction}, Phys. Rev. Lett. 111 (2013), 054802.
%
\bibitem{Green}
D.G. Green and C.N. Harvey, {\itshape Dynamics of an electron in a high intensity laser field}, arXiv:1307.8317 (2013).
%
\bibitem{DiPiazza_Review}
A. Di Piazza, C. M\"{u}ller, K.Z. Hatsagortsyan and C.H. Keitel, {\itshape Extreme high-intensity laser interactions with fundamental quantum systems}, Rev. Mod. Phys. 84 (2012), pp. 1177--1228.
%
\bibitem{Welton}
T.A. Welton, {\em Some Observable Effects of the Quantum-Mechanical Fluctuations of the Electromagnetic Field}, Phys. Rev. 74 (1948), pp. 1157--1167.
%
\bibitem{FLO1}
G.W. Ford, J.T. Lewis and R.F. O'Connell, {\em Quantum Langevin equation}, Phys. Rev. Lett. {\bf 55} (1985), pp. 2273--2276.
%
\bibitem{FLO2}
G.W. Ford, J.T. Lewis and R.F. O'Connell, {\em Quantum Oscillator in a Blackbody Radiation Field}, Phys. Rev. A {\bf 37} (1988), pp. 4419--4428.
%
\bibitem{AKE}
J.R Ackerhalt, P.L. Knight and J.H. Eberly, {\em Radiation Reaction and Radiative Frequency Shifts}, Phys. Rev. Lett. {\bf 30} (1973), pp. 456--460.
%
\bibitem{MAS}
P.W. Milonni, J.R Ackerhalt and W.A. Smith, {\em Interpretation of Radiative Corrections in Spontaneous Emission}, Phys. Rev. Lett. {\bf 31} (1973), pp. 950--960.
%
\bibitem{Heinzl:Thomson} T. Heinzl and A. Ilderton, {\em Corrections to Laser Electron Thomson Scattering}, arXiv:1307.0406 (2013).
%
\bibitem{DiPiazza:Landau-Lifshitz}
A. Di Piazza, {\em Exact Solution of the Landau-Lifshitz Equation in a Plane Wave}, Lett. Math. Phys. 83 (2008), pp. 305--313.
%
\bibitem{Hadad}
Y. Hadad {\it et al.}, {\em Effects of radiation reaction in relativistic laser acceleration}, Phys. Rev. D 82 (2010), 096012.
%
\bibitem{Lehmann:gain}
G. Lehmann and K.H. Spatschek, {\em Energy gain of an electron by a laser pulse in the presence of radiation reaction}, Phys. Rev. E 84 (2011), 046409.
%
\bibitem{Harvey:symmetry}
C. Harvey, T. Heinzl and M. Marklund, {\em Symmetry breaking from radiation reaction in ultra-intense laser fields}, Phys. Rev. D 84 (2011), 116005.
%
\bibitem{Heinzl:a0}
T. Heinzl and A. Ilderton, {\em A Lorentz and gauge invariant measure of laser intensity}, Opt.Commun.282 (2009), pp. 1879--1883.
%
\bibitem{DiPiazza:sub} A. Di Piazza, K.Z. Hatsagortsyan and C.H. Keitel, {\em Strong Signatures of Radiation Reaction below the Radiation-Dominated Regime}, Phys. Rev. Lett. 102 (2009), 254802.
%
%
%
\end{thebibliography}
\end{document}